\documentclass[journal]{IEEEtran}

\usepackage{siunitx}
\usepackage{amsmath}
\usepackage{graphicx}

\begin{document}

\title{Temporally-multiplexed dual-frequency terahertz imaging at kilohertz frame rates}%

\author{Lucy A. Downes, C. Stuart Adams, and Kevin J. Weatherill%
\thanks{Department of Physics, Durham University, South Road, Durham, DH1 3LE, UK.}%
\thanks{lucy.downes@durham.ac.uk}}%

\maketitle

\begin{abstract}
We present a temporally-multiplexed dual-colour terahertz (THz) imaging technique using THz-to-optical conversion in atomic vapour. By rapidly alternating the pump laser frequency, we sequentially excite two atomic states, each absorbing a different THz frequency: 0.5 THz and 1.1 THz. Each THz field induces optical fluorescence at a distinct wavelength, enabling the creation of a sequence of alternating, interleaved images for each frequency. Synchronizing the laser switching with camera acquisition allows video capture at 1,000 frames per second for both frequencies. The system’s speed is limited only by laser power and fibre switching hardware. The presented method can be scaled to image more THz frequencies through the addition of further laser frequencies, paving the way for THz hyperspectral imaging in many real-world settings.
\end{abstract}


\section{Introduction}

\IEEEPARstart{H}{yperspectral} and multispectral imaging involve collecting both spatial and spectral information from a target or sample. These techniques have a wide range of applications, spanning fields such as food quality control~\cite{NICOLAI200799}, environmental monitoring~\cite{Stuart2019}, security screening~\cite{Security}, and medical imaging~\cite{medical-hyperspectral}. Most multispectral imaging systems operate across the visible and near-infrared (NIR) regions, where they are used to distinguish materials that may appear identical at a single wavelength.

Imaging in the terahertz (THz) frequency range complements optical and infrared techniques, offering additional information as THz radiation can deeply penetrate many materials that are opaque to visible light and infrared, allowing sub-surface structures to be observed, while also being non-ionizing and thus safe for biological tissues~\cite{Mittleman18,Leitenstorfer_2023}. Furthermore, many materials display molecular resonances in the THz range, making chemical identification possible. These characteristics have have allowed THz multi/hyper-spectral imaging to be suggested for applications in agriculture~\cite{Ge2021}, medical imaging~\cite{Yang2016, Vohra2022}, archaeology~\cite{Dong2022} and the identification of illicit drugs~\cite{Kawase:03}.
Various THz imaging methods exist, tailored to specific applications~\cite{THzImagingRoadmap21}. For example, in applications requiring high-resolution spectral information, time-domain spectroscopy (TDS) is typically used as it can provide continuous wide-band response across the THz range~\cite{TDS_tutorial}. However TDS image acquisition is relatively slow as it requires a 2D image to be constructed pixel by pixel.
Conversely, in applications requiring higher-speed THz image acquisition, focal plane array (FPA) detectors such as microbolometer~\cite{Lee:05} or FET arrays~\cite{Bauer2014} are deployed, as they can capture 2D images in a single shot. However, spectral information is typically not collected, and these detectors struggle to capture real-time images.
Hybrid approaches also exist, where tunable THz sources, or multiple THz sources at different frequencies are used in combination with FPA detectors~\cite{Zhou2018}. 

In our previous work we introduced a technique which uses laser-excited thermal atomic vapour to convert THz frequency radiation into optical radiation that can then be imaged using standard CCD detectors~\cite{Wade2017}, offering very high speed image acquisition at kilohertz frame rates~\cite{Downes2020, Downes2023}. 
Here we demonstrate that this atom-based imaging technique is able to operate at two THz frequencies almost simultaneously, allowing high-speed videos to be captured at both frequencies. We demonstrate the speed of this technique and present proof-of-concept experiments on materials discrimination. This paves the way for a high-speed atom-based hyperspectral or multicolour THz imaging system.

\section{Methods}

\begin{figure}[ht]
\centering\includegraphics[width=0.5\textwidth]{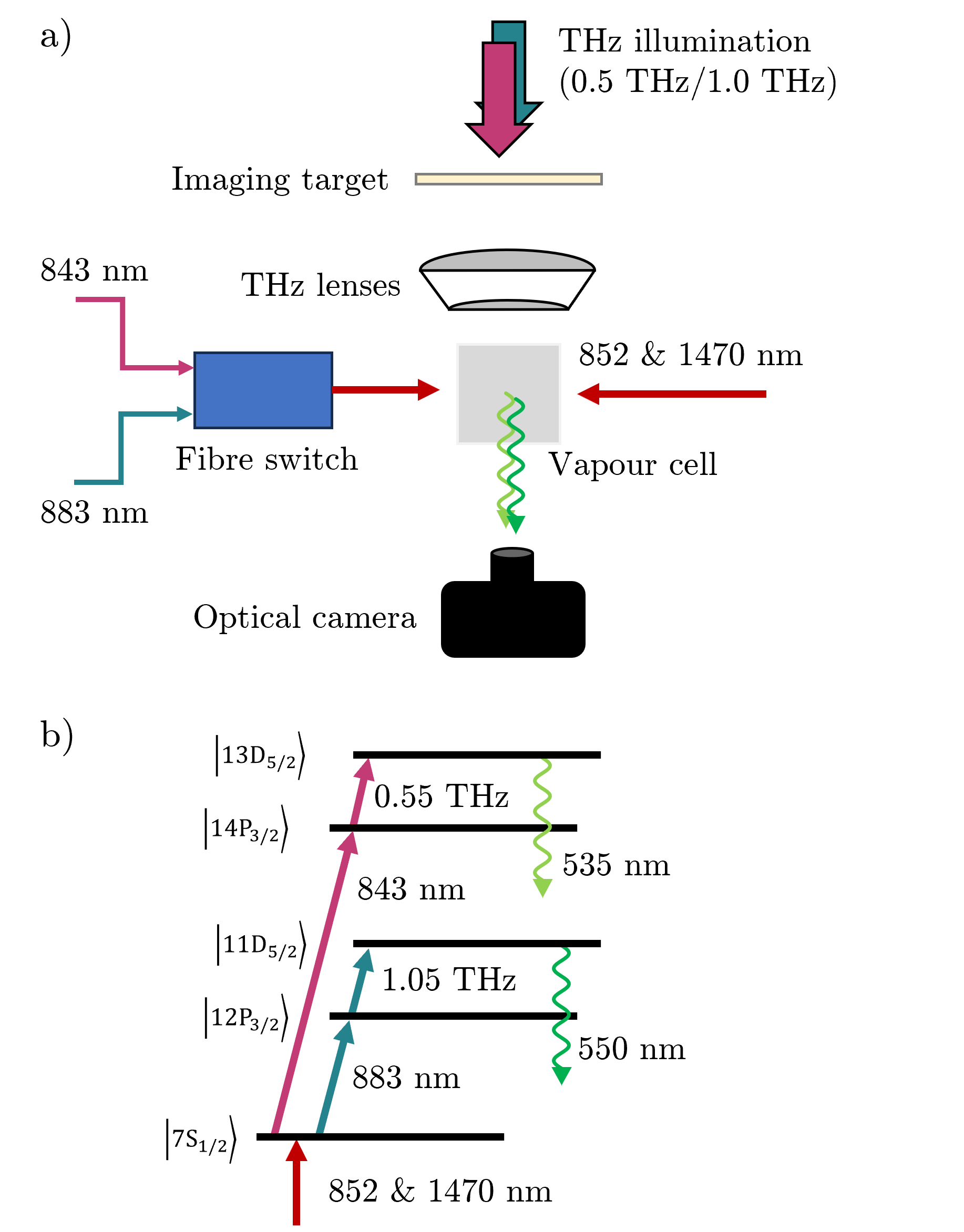}
\caption{\textbf{Experimental layout and atomic energy levels} a) A simplified illustration of the experimental setup used in this work. 3 IR laser beams are incident in a vapour cell filled with Cs, the frequency of the final laser can be selected using a fibre switch. The THz field passes through the imaging target and is incident on the vapour cell perpendicular to the direction of propagation of the lasers. The resulting optical fluorescence is then imaged on a camera. b) The atomic energy levels that are populated during the process and give rise to visible fluorescence. The lower energy levels addressed by the first two IR pump lasers are omitted for clarity.}
\label{Fig:layout}
\end{figure}

This work uses the same basic principles of atom-based THz imaging as described in previous work \cite{Downes2020, Downes_Thesis, Downes2023}, namely a 3-photon excitation scheme in a warm Caesium atomic vapour to reach a high-lying Rydberg state which emits optical fluorescence. A resonant THz field can then couple to a nearby Rydberg state causing the colour of the optical fluorescence to change. By isolating and imaging only the fluorescence from the THz-coupled state, an image of the incident THz field is formed. Since there are Rydberg transitions that span the THz frequency band, this technique can be used to image using a wide range of THz frequencies~\cite{Sibalic2017,Downes2023}. 
A diagram of the basic layout of the imaging system is shown in Figure~\ref{Fig:layout}(a). Infrared lasers are overlapped in a glass cell filled with Cs vapour such that their region of overlap creates a plane of excited atoms that are sensitive to the incoming THz field. The THz field passes through the sample to be imaged and is then incident normal to the plane of excited atoms. The visible fluorescence emitted is then captured on a camera.
In this work we image objects using fields at 0.549\,\si{\tera\hertz} and 1.055\,\si{\tera\hertz} which correspond to the $14\rm{P}_{3/2} \rightarrow 13\rm{D}_{5/2}$ and $12\rm{P}_{3/2} \rightarrow 11\rm{D}_{5/2}$ transitions respectively. 

The first two infrared (IR) pump lasers at 852\,\si{\nano\metre} and 1470\,\si{\nano\metre} are identical to previous works, and are frequency stabilised to the atomic transitions they address with powers of approximately 5\,\si{\milli\watt} and 10\,\si{\milli\watt} respectively. 

In contrast to previous experiments we now have two separate lasers for excitation to the two different initial Rydberg states; one at 843\,\si{\nano\metre} (45\,\si{\milli\watt}) to address the $7\rm{S}_{1/2}\rightarrow 14\rm{P}_{3/2}$ transition and another at 883\,\si{\nano\metre} (50\,\si{\milli\watt}) for the $7\rm{S}_{1/2}\rightarrow 12\rm{P}_{3/2}$ transition. The transitions addressed by the two pairs of laser and THz fields are shown in Figure~\ref{Fig:layout}(b). Neither of the final excitation lasers is actively frequency stabilised as they are both sufficiently passively stable over the timescales required for imaging, but they could easily be stabilised for example to an optical cavity using Pound-Drever-Hall (PDH) locking, or by using a scanning transfer cavity lock \cite{STCL,STCL2}. 
Both beams are coupled into an optical switch module (Thorlabs OSW12) such that either can be directed into the vapour cell at any time. The switching time of the optical switch was measured to be $0.2\,\si{\milli\second}$.

\begin{figure}[t]
\centering\includegraphics[width=0.5\textwidth]{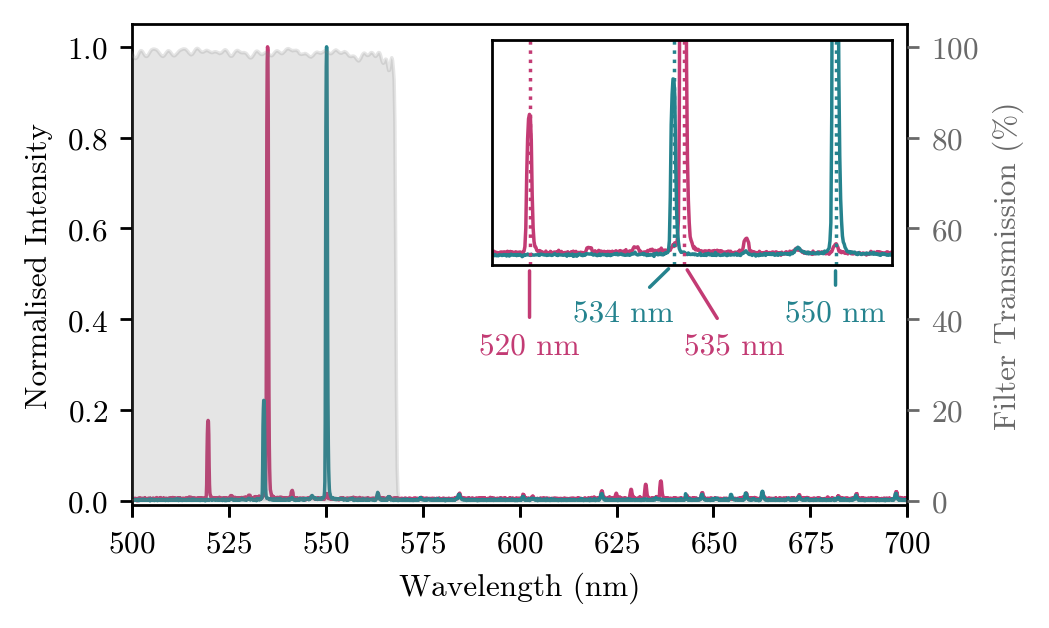}
\caption{\textbf{Fluorescence spectra and filter transmission.} The spectrum of visible fluorescence emitted by the atomic vapour in the presence of the 0.5\,\si{\tera\hertz} (pink) and 1.1\,\si{\tera\hertz} (teal) fields respectively. Both spectra are normalised such that the highest peak intensity is 1. The largest peaks in both spectra are from decay on the $n\rm{D}_{5/2}\rightarrow 6\rm{P}_{3/2}$ transitions at 535\,\si{\nano\metre} and 550\,\si{\nano\metre}, while the smaller peaks at 520\,\si{\nano\metre} and 534\,\si{\nano\metre} are from decay between lower fine structure states. The grey shaded region represents the transmission of the optical filter used for the false-colour images.}
\label{Fig:spectra}
\end{figure}

The THz fields are created using amplifier-multiplier chains (AMCs) from Virginia Diodes, seeded from a Windfreak SynthHD. The two beams are collimated using Teflon lenses and combined using a High-Resistivity Float Zone Silicon (HRFZ Si) beamsplitter with a transmission/reflection ratio of approximately 55:45. While this means that around 50\% of the power from both THz sources is lost, the atomic system is sensitive enough for this not to be an issue~\cite{Downes2020}. After considering losses from the beamsplitter and the internal attenuation settings of the sources we estimate the maximum THz power at the cell position is $150~\si{\micro\watt}$ at 1.1\,\si{\tera\hertz} and $50\si{\micro\watt}$ at 0.5\,\si{\tera\hertz}. Since the dipole matrix element of the transition driven by the 0.5\,\si{\tera\hertz} field is larger than that addressed by the 1.1\,\si{\tera\hertz} field, the difference in applied power leads to a similar signal at both frequencies. 
The THz field is then imaged onto the vapour cell using a custom THz lens system from i2S which was shown to be achromatic at our frequencies of interest.
The different fluorescence spectra emitted from the two final Rydberg states is shown in Fig.~\ref{Fig:spectra}. The emission in the presence of the 843\,\si{\nano\metre} laser and 0.5\,\si{\tera\hertz} field shows peaks at 535\,\si{\nano\metre} and 520\,\si{\nano\metre} (corresponding to decays on the $13\rm{D} \rightarrow 6\rm{P}$ transitions) whereas in the presence of the 883\,\si{\nano\metre} and 1.1\,\si{\tera\hertz} fields the fluorescence has peaks at 550\,\si{\nano\metre} and 534\,\si{\nano\metre} (corresponding to decays on the $11\rm{D} \rightarrow 6\rm{P}$ transitions). 
There was no measurable effect of the 0.5\,\si{\tera\hertz} field when imaging on the 1.1\,\si{\tera\hertz} transition and vice versa, hence both THz sources were applied simultaneously and the fibre switch used to select the THz frequency of interest.

Any optical camera can be used to image the vapour, depending on the desired characteristics of the images. The true-colour images were taken on a mobile phone camera (Samsung Galaxy S10e) through a low-pass IR filter (Thorlabs FESH780) to block out any scattered laser light. 
For longer exposure, low-noise images an Andor iXon Ultra camera was used to image the fluorescence. For these images both the camera and the optical switch were triggered by synchronised external function generators. 
To capture frames at high speeds, a Phantom VEO camera was used. In this case, the optical switch was triggered from a pulse output by the camera which set the frame rate for the acquisition. Due to the non-zero response time of the optical switch, the delay of this output pulse had to be optimised to eliminate crosstalk between frames. 
Again, an optical filter (Semrock FF01-505/119) was used to block out unwanted background light. The transmission profile of this filter is illustrated by the shaded region in Fig.~\ref{Fig:spectra}.

\section{Results}

\begin{figure}[t]
\centering\includegraphics[width=0.48\textwidth]{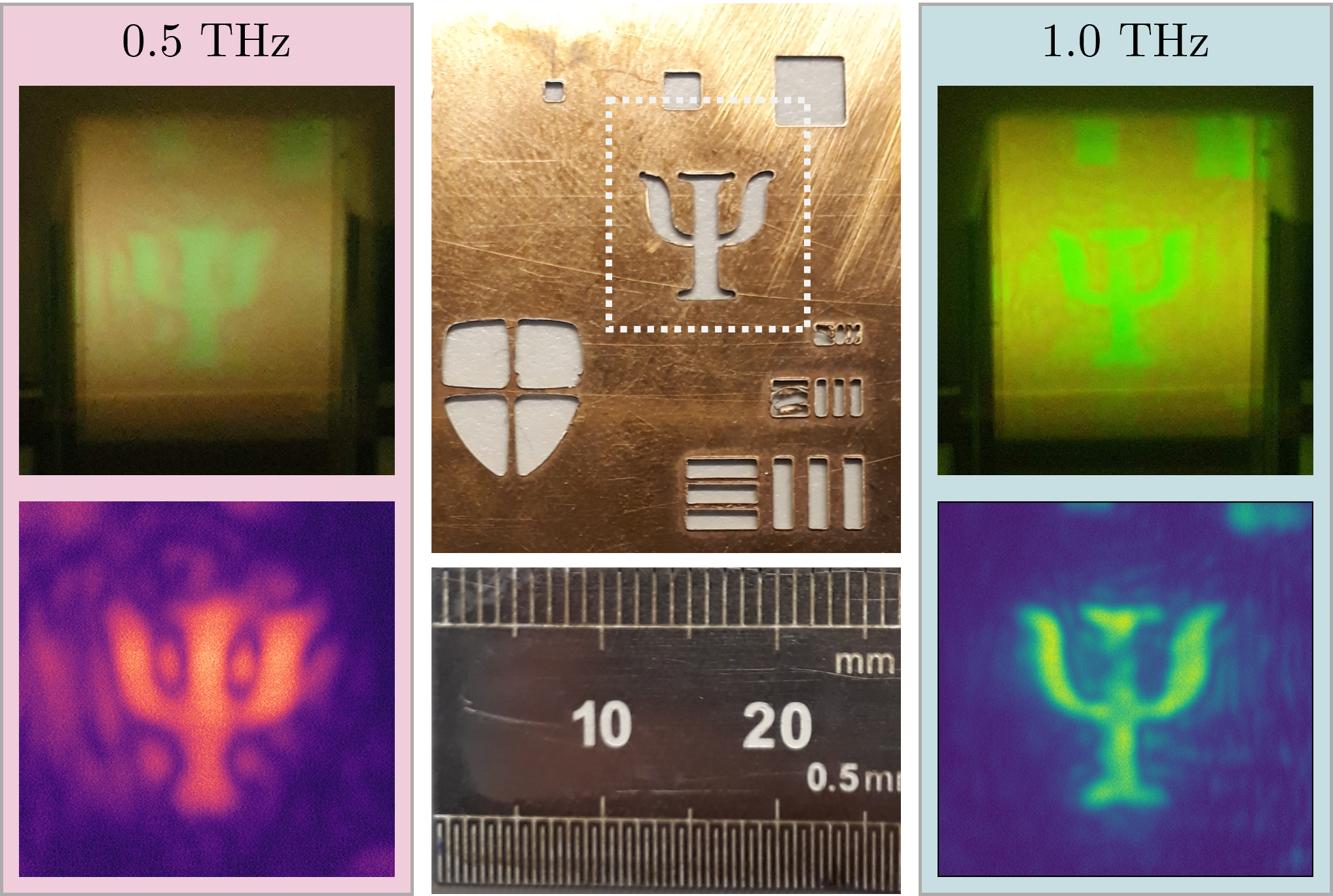}
\caption{\textbf{Example images at 0.5 and 1.1\,\si{\tera\hertz}.} True colour (\textit{top}) and false-colour background-subtracted (\textit{bottom}) terahertz images of a `Psi'-shaped aperture in a metal mask (\textit{centre}). The images on the right (taken at 1.1\,\si{\tera\hertz}) show increased spatial resolution compared to those on the left (taken at 0.5\,\si{\tera\hertz}). The mask is identical to the one used in \cite{Downes2020}.}
\label{Fig:Psis}
\end{figure}

Figure \ref{Fig:Psis} shows examples of images taken at both 0.5\,\si{\tera\hertz} and 1.1\,\si{\tera\hertz}. A thin metal mask with a `Psi'-shaped aperture is placed at the focus of the THz imaging system (position marked `imaging target' in Fig.~\ref{Fig:layout}(a)) and the resulting fluorescence is imaged with both a colour camera (top) and high-resolution monochrome camera (bottom). The position of the mask, THz lenses and cameras remained the same between each image, only the final laser frequency was varied to image the two THz fields. There is a clear improvement in spatial resolution between the images at 0.5\,\si{\tera\hertz} (left) and those at 1.1\,\si{\tera\hertz} (right), due to the reduction in wavelength by a factor of 2. The true-colour images show the slight difference in the colour of the fluorescence emitted from the different atomic states addressed. For the high resolution images, an image of the vapour in the absence of the THz fields has been subtracted to remove any unwanted background fluorescence. Here the exposure time was 200\,\si{\milli\second}.

\begin{figure}[t]
\centering\includegraphics[width=0.48\textwidth]{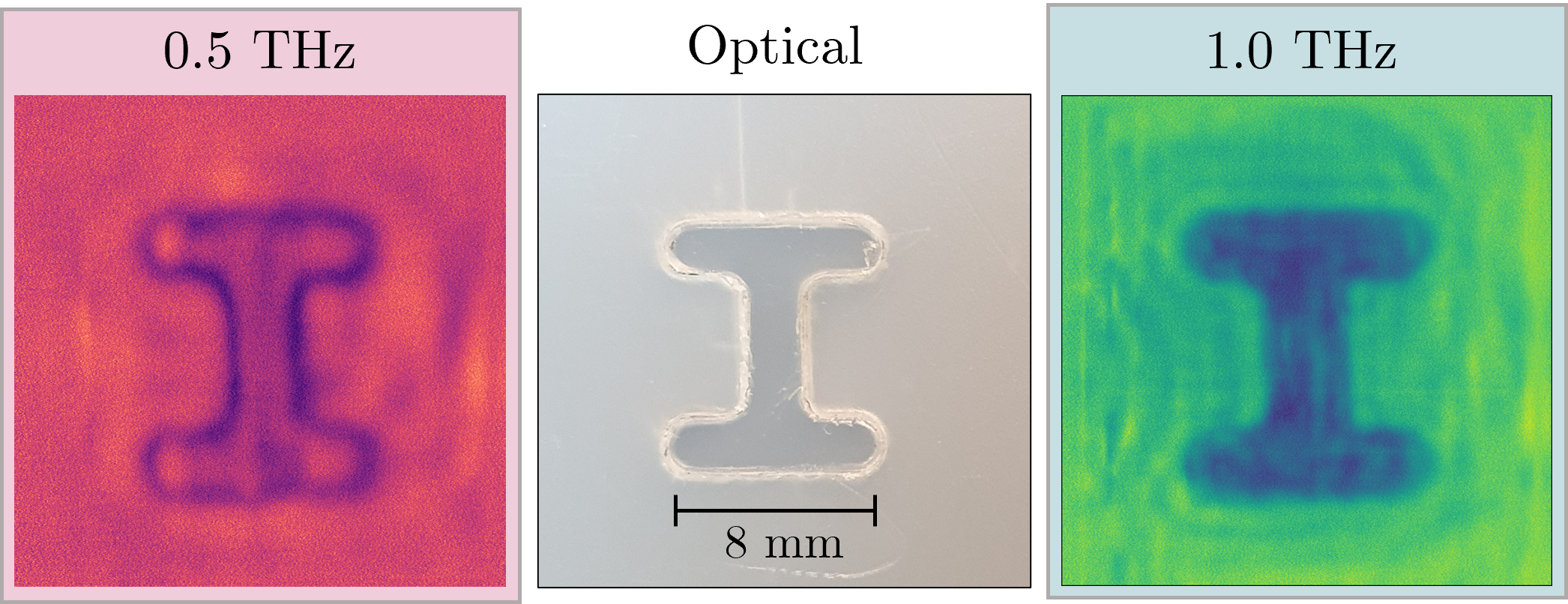}
\caption{\textbf{Material discrimination using THz imaging.} Images of a letter I-shaped target made from Nylon (inner section) and polypropylene (outer section). At optical frequencies (\textit{centre}) the plastics appear visually similar. At 0.5\,\si{\tera\hertz} (\textit{left}), both materials exhibit similar transmission whereas at 1.1\,\si{\tera\hertz} (\textit{right}) the nylon becomes opaque.}
\label{Fig:materials}
\end{figure}

As a proof-of-principle demonstration of how this technique can be used to distinguish between materials we image a sample composed of two optically similar materials at two THz frequencies, as shown in Fig.~\ref{Fig:materials}. 
To create the composite sample, a shape is cut out from a thin ($1\,\rm{mm}$ thickness) polypropylene sheet and a similarly-shaped thin ($1\,\rm{mm}$ thickness) nylon piece is inserted into the gap. At visible wavelengths the two plastics have very similar colour and opacity making them hard to distinguish (Fig.~\ref{Fig:materials}, centre). When imaged at 0.5\,\si{\tera\hertz} (Fig.~\ref{Fig:materials}, left) the polypropylene and nylon show similarly low absorption, however at 1.1\,\si{\tera\hertz} (Fig.~\ref{Fig:materials}, right) the nylon is almost opaque whereas the polypropylene is still transparent allowing them to be differentiated. 
While both materials are transparent at 0.5\,\si{\tera\hertz}, the edges of the shape are clearly visible in the image. The method used to cut the plastics left rough edges, as can been seen in the optical image, meaning that the materials are marginally thicker at this interface leading to increased THz absorption. The increased surface roughness also leads to scattering at the material interface, enhancing the appearance of the edges. Again the exposure time for the THz images is 200\,\si{\milli\second} and an image of the fluorescence without the THz field has been subtracted to remove any unwanted background.

\begin{figure*}[ht]
\centering\includegraphics[width=\textwidth]{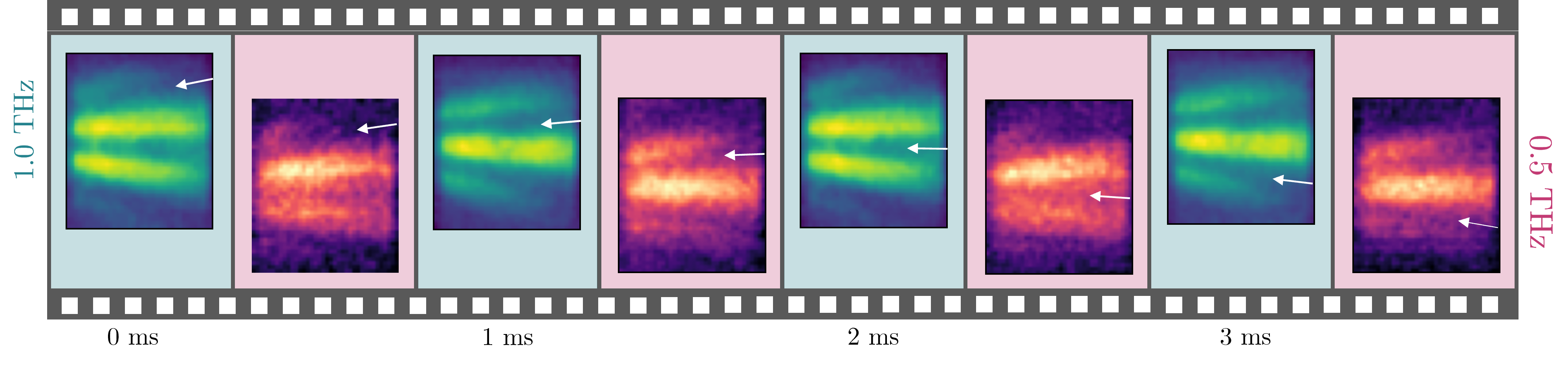}
\caption{\textbf{High speed two-colour imaging} Sequential frames taken at 2000~fps of an optical chopper wheel rotating at 500~rpm. The THz frequency used alternates between images; the teal background shows frames imaged at 1.1\,\si{\tera\hertz} while the pink background uses 0.5\,\si{\tera\hertz}. The white arrow highlights the motion of a single spoke of the wheel between frames. Binning, read-noise subtraction and smoothing has been applied to each frame. Example videos can be found in the supplementary information.}
\label{Fig:highspeed}
\end{figure*}

To demonstrate how this technique is able to capture videos at two different THz frequencies effectively simultaneously, we image a chopper wheel rotating at 500 rpm. During the acquisition, both THz fields are continually incident on the vapour while the optical switch is used to alternate between the laser fields at a rate matching the frame-rate of the camera. In this case, the camera frame-rate was 2000 frames per second, resulting in two videos each at 1000 fps. Fig.~\ref{Fig:highspeed} shows subsequent frames from the acquisition which correspond to images at 0.5\,\si{\tera\hertz} and 1.1\,\si{\tera\hertz} respectively. 

After the acquisition is complete the frames can be demultiplexed and processed before being reassembled into separate videos. The required post-processing is minimal, and here consists of binning ($8\times8$), read-noise subtraction and smoothing using a uniform filter to enhance clarity.
Again the frames taken at 1.1\,\si{\tera\hertz} (teal background) show higher resolution than those taken at 0.5\,\si{\tera\hertz} (pink background). 

\section{Discussion}

Since the response of the Rydberg atoms is narrowband the presented scheme has excellent out-of-band rejection, meaning that there is no crosstalk between illumination frequencies. This eliminates the need to tune the THz between imaging frequencies and enables high-speed image capture. The presented methodology could be extended to imaging at more THz frequencies through the addition of more laser wavelengths to address more Rydberg transitions. This would require multiple optical switches to multiplex the additional laser fields. Atom-based systems can also offer S.I. traceable electric field measurements in the THz band, eliminating the need for external calibration at each frequency~\cite{Chen:22}.

The difference in signal-to-noise between images at the two frequencies is due to the amount of fluorescence emitted from each atomic state. This fluorescence is related to the amount of laser power available in the final excitation step, as well as properties of the atomic state used~\cite{Downes2023}. While earlier we noted that the fluorescence levels were similar, we are limited by the power in the 843\,\si{\nano\metre} laser ($<45\,\si{\milli\watt}$ compared to $>50\,\si{\milli\watt}$ at 883\,\si{\nano\metre}) and hence the level of fluorescence from the 0.5\,\si{\tera\hertz} transition is lower than that from the transition addressed by the 1.1\,\si{\tera\hertz} field. This is particularly evident at very short exposures/high frame-rates and is the reason for the increase in noise in the 0.5\,\si{\tera\hertz} images in Fig.~\ref{Fig:highspeed}. This could be easily improved by increasing the power available at both laser wavelengths.

The atomic states chosen in this work were selected as they have previously been shown to be suitable for imaging with our currently available THz source frequencies~\cite{Downes2023, Sibalic2017}, however the fluorescence spectra `overlap' as seen in Fig.~\ref{Fig:spectra} and so cannot be spectrally separated without significant crosstalk or losses. This means that true simultaneous imaging is not possible using these states. Selecting transition pairs that are separated by a larger frequency, for example $11\rm{P}_{3/2}\rightarrow10\rm{D}_{5/2}$ (at 1.5\,\si{\tera\hertz}) and $14\rm{P}_{3/2}\rightarrow13\rm{D}_{5/2}$ (at 0.5\,\si{\tera\hertz}), would enable simultaneous imaging.

An alternative method for true simultaneous multispectral THz imaging would be to use two different atomic species to image different THz frequencies, for example Rb and Cs. The fluorescence emitted by THz-coupled states in Rb is blue-shifted compared to Cs and so could be easily isolated using readily available filters.

\section{Conclusion}

We have demonstrated temporally-multiplexed dual-frequency terahertz imaging at kilohertz frame rates using atomic vapour. By switching rapidly between two different excitation laser wavelengths we are able to create high-speed videos at two distinct THz frequencies with no crosstalk. This will pave the way for materials discrimination using THz hyperspectral imaging in many real-world settings. 

\section*{Acknowledgements}

We gratefully acknowledge funding by the UK Engineering and Physical Sciences Council via grants No. EP/MO1326X/1, No. EP/W033054/1, No. EP/V030280/1, and No. EP/W009404/1.

\bibliographystyle{IEEEtran}

\bibliography{THz}
\end{document}